\documentclass[reprint,superscriptaddress, amsmath,amssymb, aps, prb,]{revtex4-1}

\usepackage{graphicx}
\usepackage{dcolumn}
\usepackage{bm}
\usepackage{hyperref}
\usepackage[mathlines]{lineno}
\usepackage{siunitx}
\usepackage{xspace}
\usepackage{epstopdf}
\usepackage{titlesec}
\titlespacing\subsubsection{0pt}{12pt plus 4pt minus 2pt}{6pt plus 2pt minus 2pt}
\usepackage{multirow}
\usepackage[normalem]{ulem}

\usepackage{enumitem}
\setlist{nosep}
\newlength{\mylen}
\setbox1=\hbox{$\bullet$}\setbox2=\hbox{\tiny$\bullet$}
\usepackage[]{natbib}

\usepackage[caption=false]{subfig}

\newcommand{\LAO}{$\mathrm{LaAlO_3}$ \xspace}
\newcommand{\STO}{$\mathrm{SrTiO_3}$ \xspace}
\newcommand{\LAOSTO}{$\mathrm{LaAlO_3}$-$\mathrm{SrTiO_3}$ \xspace}

\usepackage{color} 

\usepackage{hyperref}
\hypersetup{
colorlinks=true,final=true,
        linkcolor=blue,
        citecolor=blue,
        filecolor=blue,
        urlcolor=blue,
}

\begin{document}

\preprint{APS/123-QED}

\title{In-gap features in superconducting \texorpdfstring{$\mathrm{LaAlO_3}$-$\mathrm{SrTiO_3}$} \quad \-  interfaces observed by tunneling spectroscopy}

\author{Lukas Kuerten}
 \email{l.kuerten@fkf.mpg.de}
\affiliation{%
Max-Planck-Institute for Solid State Research,\\
70569 Stuttgart, Germany
}%
\author{Christoph Richter}
\author{Narayan Mohanta}
\author{Thilo Kopp}
\author{Arno Kampf}
\affiliation{%
Center for Electronic Correlations and Magnetism, \\
Theoretical Physics III and Experimental Physics VI, \\
Institute of Physics, University of Augsburg, 86135 Augsburg, Germany 
}%
\author{Jochen Mannhart}
\author{Hans Boschker}%
 \email{h.boschker@fkf.mpg.de}
\affiliation{%
Max-Planck-Institute for Solid State Research,\\
70569 Stuttgart, Germany
}%

%
%

\date{\today}

\begin{abstract}
We identified quasiparticle states at well-defined energies inside the superconducting gap of the electron system at the \LAOSTO interface using tunneling spectroscopy. The states are found only in a number of samples and depend upon the thermal-cycling history of the samples. The states consist of a peak at zero energy and other peaks at finite energies, symmetrically placed around zero energy. These peaks disappear, together with the superconducting gap, with increasing temperature and magnetic field. We discuss the likelihood of  various physical mechanisms that are known to cause in-gap states in superconductors and conclude that none of these mechanisms can easily explain the results. The conceivable scenarios are the formation of Majorana bound states, Andreev bound states, or the presence of an odd-frequency spin triplet component in the superconducting order parameter. 

\end{abstract}

\maketitle


\section{Introduction}

Superconductors are characterized by the opening of a gap in the quasiparticle spectrum at the Fermi energy. The presence of states inside this gap indicates physics beyond conventional superconducting behavior and is, therefore, an exciting topic in science~\cite{Hall1960, Rowell1966, Altshuler1979,Buchholtz1981, Blonder1982, Alff1998,Kashiwaya1995,  Kashiwaya2000,Lofwander2001, Deutscher2005, SanGiorgio2008, Linder2010, Boden2011, diBernardo2015, Kitaev2001, Fu2008, Alicea2012, Mourik2012, Beenakker2013, Yu1965,Shiba1968,Rusinov1969,Zittartz1970, Kirtley2007}. There are different mechanisms that can cause a finite spectral density inside the superconducting gap. For example, for nodal superconductors, only a part of the Fermi surface is gapped, resulting in a smooth variation of the density of quasi-particle states as a function of energy inside the gap.
 In some cases, however, a peak in the spectral density is present at zero energy, or multiple peaks are present at finite energies. These peaks can be caused by, for example, Andreev bound states at interfaces between unconventional superconductors and normal metals~\cite{Buchholtz1981, Blonder1982, Alff1998,Kashiwaya1995,  Kashiwaya2000,Lofwander2001, Deutscher2005},  an odd-frequency spin triplet component of the superconducting order parameter~\cite{SanGiorgio2008, Linder2010, Boden2011, diBernardo2015}, the solid-state analog of Majorana fermions~\cite{Kitaev2001, Fu2008, Alicea2012, Mourik2012, Beenakker2013}, and by bound states due to the presence of magnetic impurities~\cite{Yu1965,Shiba1968,Rusinov1969,Zittartz1970}. Zero bias anomalies also frequently appear in tunneling studies on high-temperature cuprate superconductors \cite{Kirtley2007}.The study of the in-gap states gives crucial information about the pairing symmetry of a superconductor. Here we report the presence of quasiparticle states inside the superconducting gap of the two-dimensional \LAOSTO interface superconductor.

At the interface of the two insulators \STO and $\mathrm{LaAlO_3}$, a conducting two-dimensional electron  system (2DES) with fascinating properties exists \cite{Ohtomo2004}. In contrast to the more conventional 2DESs, which exist, e.g., at semiconductor heterointerfaces where the electrons are dilute and behave like a free electron gas, the conduction electrons at the \LAOSTO interface stem from local Ti 3d orbitals and  exhibit unusual and novel properties. Due to correlations, the 2DES is often referred to as a two-dimensional electron liquid (2DEL)~\cite{Breitschaft2010}.  To name only a few of these  properties, the interface is gate-tunable \cite{Thiel2006} and exhibits superconductivity \cite{Reyren2007}, which is also gate-tunable~\cite{Caviglia2008}. In addition, it is reported to be a host to a large number of other interesting phenomena, such as (gate-tunable) Rashba spin-orbit coupling~\cite{Caviglia2010, BenShalom2010, Zhong2013} and the coexistence of superconductivity and magnetism \cite{Li2011, Bert2011}. More aspects of the $\mathrm{LaAlO_3}$-$\mathrm{SrTiO_3}$-interface 2DEL can be found in several review articles~\cite{Hwang2012, Gariglio2016, Boschker2017}.

Recently, we performed tunneling measurements on \LAOSTO interfaces, allowing us to measure the superconducting gap, map the corresponding phase diagram \cite{Richter2013, Fillis-Tsirakis2016}, and to identify electron-phonon coupling as a likely origin of the superconductivity \cite{Boschker2015}. Almost all \LAOSTO tunneling samples investigated exhibit superconducting gap spectra with the expected BCS density of states consisting of a full gap and coherence peaks. In some cases, however, we observed spectra which exhibit distinct peaks inside the superconducting gap. The in-gap features appear and disappear non-deterministically upon different thermal cycles and gate-voltage sweeps. In this manuscript, we describe the structure and occurrence of these states and discuss the most likely scenarios of their origin.

\section{In-gap states}
\label{sec:In-gap_states} 

We give an overview over a selection of various in-gap states that can be observed in superconducting tunnel junctions and briefly explain their origins and properties. These states are summarized in Table \ref{tab:explanations}.

\setlength{\tabcolsep}{10pt}
\renewcommand{\arraystretch}{1.75}

\begin{table*}
\caption{A compilation of mechanisms that can induce in-gap states in tunnel spectra of superconductors (SCs). For more detailed
information and additional references, see Sec. II. The left column
denotes the names and origins of the phenomena and the right column
describes their properties. Note that many of these states are interdependent. Their relation to one another is further discussed in Sec II.}
\begin{tabular}{m{0.35\textwidth} |  m{0.6\textwidth}}

\hline

\textbf{Name} &
\multirow{2}{0.6\textwidth}{\\[-1.5ex]\textbf{Properties}} \\
\textbf{Origin} &\\

\hline

\textbf{Kondo resonance} \cite{Kondo1964} & \multirow{2}{0.6\textwidth}{
\begin{itemize}
\item Resonance effects observable as peaks in conductivity.
\item Zero-bias peak splits in a finite magnetic field.
\item Side-peak separation varies linearly with external magnetic field.
\item Does not require superconductivity
\end{itemize}
}\\
Resonance at magnetic impurities located inside the conducting host. &\\
\\
\hline

\textbf{Anderson-Appelbaum states}\cite{Anderson1966, Appelbaum1966, Appelbaum1967} &
\multirow{2}{0.6\textwidth}{
\begin{itemize}
\item Similar to Kondo resonance (see above)
\end{itemize}
}\\
Exchange interaction between tunneling electrons and magnetic impurities located inside the tunnel barrier. &\\
\hline

\textbf{Impurity states} &  \multirow{2}{0.6\textwidth}{
\begin{itemize}
\item Decrease of conductivity at zero bias (barrier states)~\cite{Giaever1968}
\item Increase of conductivity at zero bias (surface states)~\cite{Samokhin2001}.
\item In-gap states are particle-hole asymmetric.
\end{itemize}
}
\\
Tunneling via intermediate impurity states in barrier or surface.&\\
\hline

\textbf{Josephson junction characteristics} & \multirow{2}{0.6\textwidth}{
\begin{itemize}
\item Gap of size $\Delta_1 +\Delta_2$.
\item Cooper-pair tunneling DC Josephson current at zero bias.
\item Peaks inside the larger gap at the gap difference $\pm|\Delta_1-\Delta_2|$.
\end{itemize}
}\\
Tunneling from $\mathrm{SC_1}$ to $\mathrm{SC_2}$. & \\
\\
\hline

\textbf{Multiband Superconductivity}\cite{Binnig1980} & \multirow{2}{0.6\textwidth}{
\begin{itemize}
\item Two gaps inside one another.
\item Two pairs of coherence peaks.
\end{itemize}
}\\
SC pairing in multiple bands. & \\
\hline

\textbf{Caroli-de Gennes-Matricon states} \cite{Caroli1964} &
\multirow{2}{0.6\textwidth}{
\begin{itemize}
\item States below the gap energy. 
\item Bound states which are localized at the core of vortices.
\item Comparable to Andreev Bound states (see below).
\end{itemize}
} \\
Andreev reflection at a vortex core.&\\
\\
\hline

\textbf{Yu-Shiba-Rusinov states} \cite{ Yu1965,Shiba1968,Rusinov1969} &
\multirow{2}{0.6\textwidth}{
\begin{itemize}
\item Paired peaks symmetric around zero energy.
\item States are localized at the impurity sites.
\item Peak positions move  with varying magnetic field~\cite{Zittartz1970}.
\end{itemize}
}\\
Bound states due to magnetic impurities in SC. & \\

\hline

\textbf{Majorana bound states} \cite{Kitaev2001, Fu2008} &
\multirow{2}{0.6\textwidth}{
\begin{itemize}
\item Zero-energy bound state for well-separated Majoranas.
\item Paired states at finite energies for interacting Majoranas \cite{Nilsson2008, Beenakker2013, Flensberg2010}.
\item Located at defects at which the SC gap closes.
\item Conductance peak height quantized in units of $2e^2/h$ for specific situations.
\end{itemize}
} \\
Emergent states at the boundary of topological superconductors. & \\
\\
\hline
\textbf{Andreev Bound states}\cite{Deutscher2005} &
\multirow{2}{0.6\textwidth}{
\begin{itemize}
\item For non-$s$-wave NS junction: peak  at zero energy. 
\item For SNS junction: peaks at finite energies,  depending on the phase difference between the SCs.
\end{itemize}
}\\
Successive Andreev reflections at NS-interfaces. & \\
\hline

\textbf{Odd-frequency spin triplet pairing}   &
\multirow{2}{0.6\textwidth}{
\begin{itemize}
\item Peaks at zero or finite energies depending on layer thickness and disorder.  \cite{SanGiorgio2008, Linder2010, Boden2011, diBernardo2015, Eschrig2008}
\item Two pairs of coherence peaks in density of states.
\end{itemize}
}\\
Induced $p$-wave pairing at interfaces between SC and inhomogeneous ferromagnet. &\\

\hline
\end{tabular}

\label{tab:explanations}
\end{table*}

\subsubsection*{Kondo resonance}

The Kondo effect \cite{Kondo1964}  introduces a resonance at magnetic impurities, e.g., in tunneling processes through quantum dots \cite{Goldhaber-Gordon1998}. The resonance effect is observable as a peak in conductivity. The peak appears usually at zero voltage bias, but can also be generated at finite voltage bias \cite{Sellier2005}. Kondo resonances do not require superconductivity and, therefore, the in-gap states which originate from this mechanism do not in general disappear upon a superconducting transition. In the presence of an external magnetic field, the zero-bias peak splits by an amount equal to the Zeeman energy of the magnetic field. The splitting of the side peaks varies linearly with magnetic field~\cite{Costi2000}. With increasing temperature, there is a reduction and broadening of the Kondo resonance~\cite{Hanke2005}.

\subsubsection*{Anderson-Appelbaum states}
In the early 1960s, a zero-bias anomaly was observed in tunneling experiments in $p$-$n$ junctions~\cite{Hall1960,Logan1964} and in tunnel junctions composed of normal metals separated by oxide barriers~\cite{Wyatt1964}. It was found that the zero-bias conductance peak varies logarithmically with temperature and, thereafter, this zero-bias anomaly was also known, in the literature, as the logarithmic anomaly. Anderson~\cite{Anderson1966} and Appelbaum~\cite{Appelbaum1966, Appelbaum1967} showed that magnetic impurities located inside the tunneling barrier close to the electrodes can participate in an exchange interaction with the tunneling electrons, resulting in the zero-bias anomaly. The characteristics of these states are similar to those generated by the Kondo resonance.

\subsubsection*{Barrier and surface impurities}
In tunnel junctions with a thick barrier, in which the probability for direct tunneling from one electrode to the other is small, impurity states enclosed in the barrier provide an alternative pathway for tunneling. In this case, electrons can tunnel via an intermediate state localized on the impurity state. Due to the finite capacitance of the particle, a minimum charging energy is required to add an electron to it. Such junctions show, therefore, a suppression in conductance around zero voltage bias \cite{Giaever1968}. Localized states due to imperfections of the surface of a superconductor may  facilitate bound states at low energy, causing peaks in the conductance around zero voltage bias \cite{Samokhin2001}.

\subsubsection*{Superconductor-insulator-superconductor (SIS) tunneling}
The tunnel spectra observed in junctions  of two superconductors separated by an insulating barrier show  gaps of  $\Delta_1 +\Delta_2$, where $\Delta_1$ and $\Delta_2$ are the gaps in the spectra  of the two superconductors.  At zero bias, a DC Cooper-pair current is observable, which depends on the relative phase of the two superconductors according to the first Josephson effect ~\cite{Josephson1962,Anderson1963}. At finite temperatures, a tunneling current flows at bias voltage equal to $\pm|\Delta_1-\Delta_2|$.

\subsubsection*{Multiband superconductivity}
If two bands of a material participate in superconductivity, tunneling spectra can reveal a double gap structure with two sets of coherence peaks corresponding to the two separate pairing strengths. The coherence peaks of the band with smaller gap width will appear inside the gap of the band with larger pairing strength \cite{Binnig1980}.  In contrast to SIS-tunneling, the double gap of a multiband superconductor does not show a Josephson-current peak at zero bias.

\subsubsection*{Caroli-de Gennes-Matricon states}
The electron system in  the core of an Abrikosov vortex in a type-II superconductor can host localized fermionic bound states which are populated at energies smaller than the superconducting gap energy \cite{Caroli1964}. The origin of this effect can be regarded as a specific form of Andreev reflection (see below), since a vortex core is a normal metal confined in a  superconductor.

\subsubsection*{Yu-Shiba-Rusinov states}
Based on  Abrikosov-Gorkov theory \cite{Abrikosov1961}, Yu \cite{ Yu1965},  Shiba \cite{Shiba1968} and Rusinov \cite{Rusinov1969} showed that magnetic impurities can facilitate bound states inside the superconducting gap. Tunneling measurements revealing such states were performed early \cite{Woolf1965}, and it was possible to resolve the contribution of individual impurity atoms on a Nb surface by scanning tunneling spectroscopy~\cite{Yazdani1997, Salkola1997, Flatte1997}. Yu-Shiba-Rusinov (YSR) states always appear in pairs symmetrically around zero energy and move to higher energy with increasing magnetic field \cite{Zittartz1970}.  For a review on impurity states in superconductors, see, e.g., Ref.~[\onlinecite{Balatsky2006}].

\subsubsection*{Majorana bound states}
Majorana fermions are particles which are their own antiparticles. While it is not yet clear whether elementary Majorana particles exist, there is growing evidence that collective states with Majorana-like properties can be created in solid-state systems \cite{Mourik2012, Nadj-Perge2014}. Superconductors are the primary candidate to host  such Majorana states, since  Bogoliubov quasiparticles, the elementary excitations of a superconductor, are particle-hole symmetric and can indeed be described using Majorana's original equations \cite{Chamon2010}. Since a pair of Majoranas constitutes a Dirac fermion, the challenge is to create unpaired or spatially separated Majoranas. The two most common proposals to achieve this goal are either to combine a superconductor with a material with strong spin-orbit coupling (e.g. a topological insulator) \cite{Fu2008, Mourik2012, Das2012} or to create a ferromagnetic chain at the surface of a superconductor \cite{Kitaev2001, Nadj-Perge2014}. In these cases, topological superconductivity is induced and isolated Majorana zero modes emerge at the edges of one-dimensional sample structures.
If the two Majorana pair partners are well-separated, Majorana Bound States (MBS) are observed at zero energy, where superconducting quasiparticles are electron-hole symmetric. However, if the wavefunctions of the two pair partners overlap, a level splitting to finite energies can occur \cite{Beenakker2013, Nilsson2008, Flensberg2010}. In the limit of zero temperature and ballistic conductance, the height of the zero-bias peak generated by  MBS is quantized in units of $2e^2/h$. Reviews of the research on MBS in solid-state and other systems can be found e.g. in Refs.~[\onlinecite{Beenakker2013, Leijnse2012, Elliott2015}].

\subsubsection*{Andreev bound states}

Andreev reflections can facilitate bound states at zero energy in normal metal--superconductor (NS) junctions if the pairing symmetry of the superconductor is non s-wave \cite{Buchholtz1981, Blonder1982, Alff1998, Deutscher2005, Lofwander2001, Kashiwaya1995, Kashiwaya2000}. Andreev bound states (ABS) also appear in SNS junctions, irrespective of the pairing symmetry \cite{deGennes1963, Tomasch1965, Rowell1966, McMillan1966}. It has been reported that ABS can also be generated at a SC-vacuum interface without any normal metal \cite{Hu1994, Carrington2001}. The ABS in SNS-junctions  is located at zero bias if there is a phase difference of $\pi$ between the two superconductors. For other phase differences, it consists of two peaks at finite energies  \cite{Blonder1982, Kashiwaya2000}. The peak height of these bound states may exceed the normal-state conductance. A zero-bias peak generated by an Andreev bound state is expected to split upon the application of magnetic field \cite{Fogelstrom1997}. Andreev processes require sufficiently good conductance between the superconductor and the normal metal, usually a NS-contact without insulating barrier.

\subsubsection*{Odd-frequency spin triplet pairing}
At superconductor--ferromagnet (SF) interfaces, a long-range proximity effect can induce odd-frequency superconductivity in the ferromagnet if magnetic disorder is present \cite{Eschrig2008}. This proximity-coupled superconductivity exhibits an odd-frequency, s-wave, spin-triplet component of the superconducting order parameter which generates in-gap peaks in tunnel spectra. The peaks appear at either finite or zero energy, depending on the relative thicknesses of the ferromagnetic and superconducting layers and on the magnetic disorder of the magnetic layer\cite{SanGiorgio2008, Linder2010, Boden2011, diBernardo2015}.
 
\vspace{12pt} 
 
Having discussed a number of effects which can cause states inside the superconducting gap in tunnel spectra, we now turn to the description of our experimental observations.

\section{Experiments}

\begin{figure}
\centering
\subfloat[]{
\includegraphics[width=0.4\textwidth]{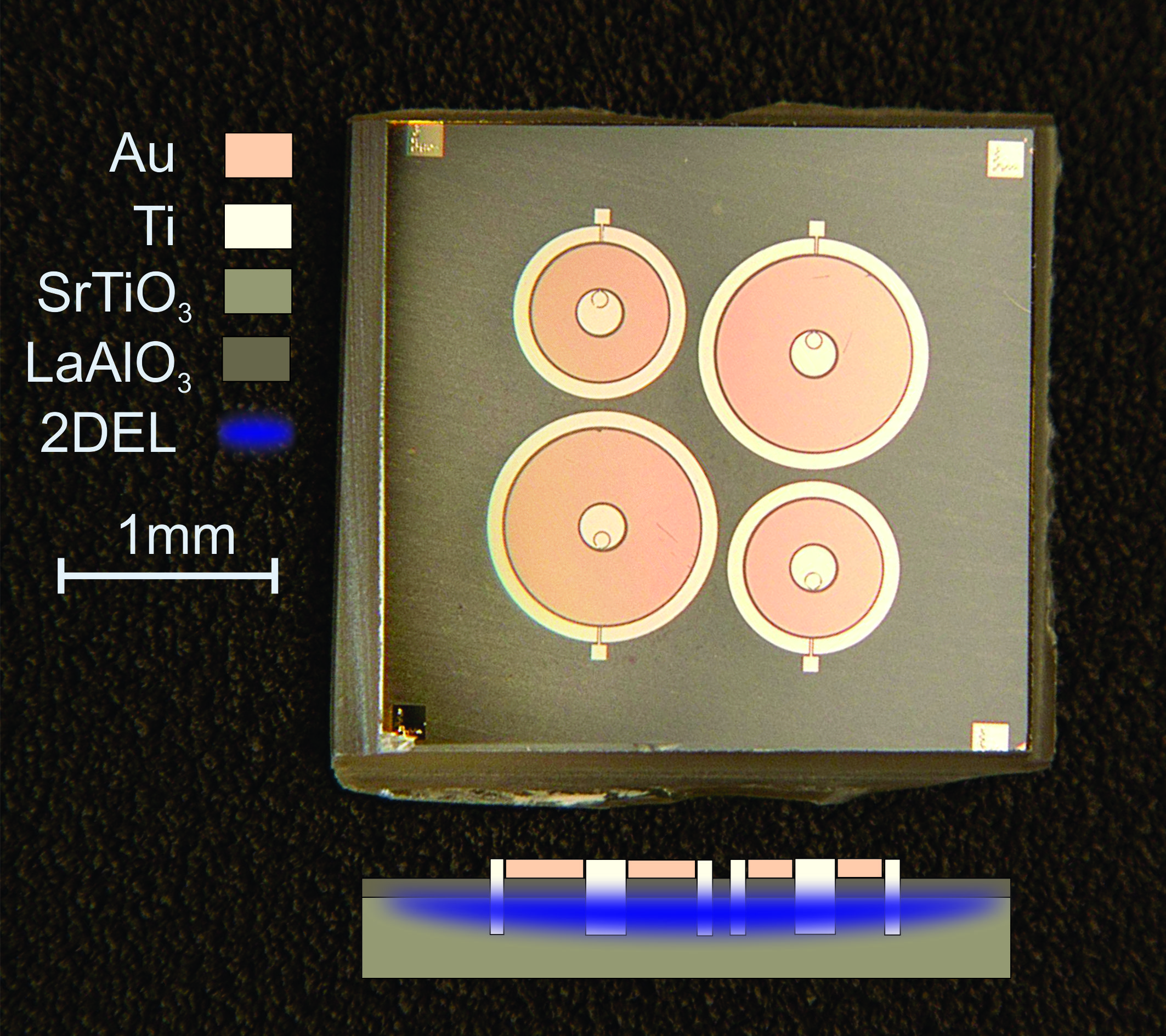}
}

\subfloat[]{
\includegraphics[width=0.4\textwidth]{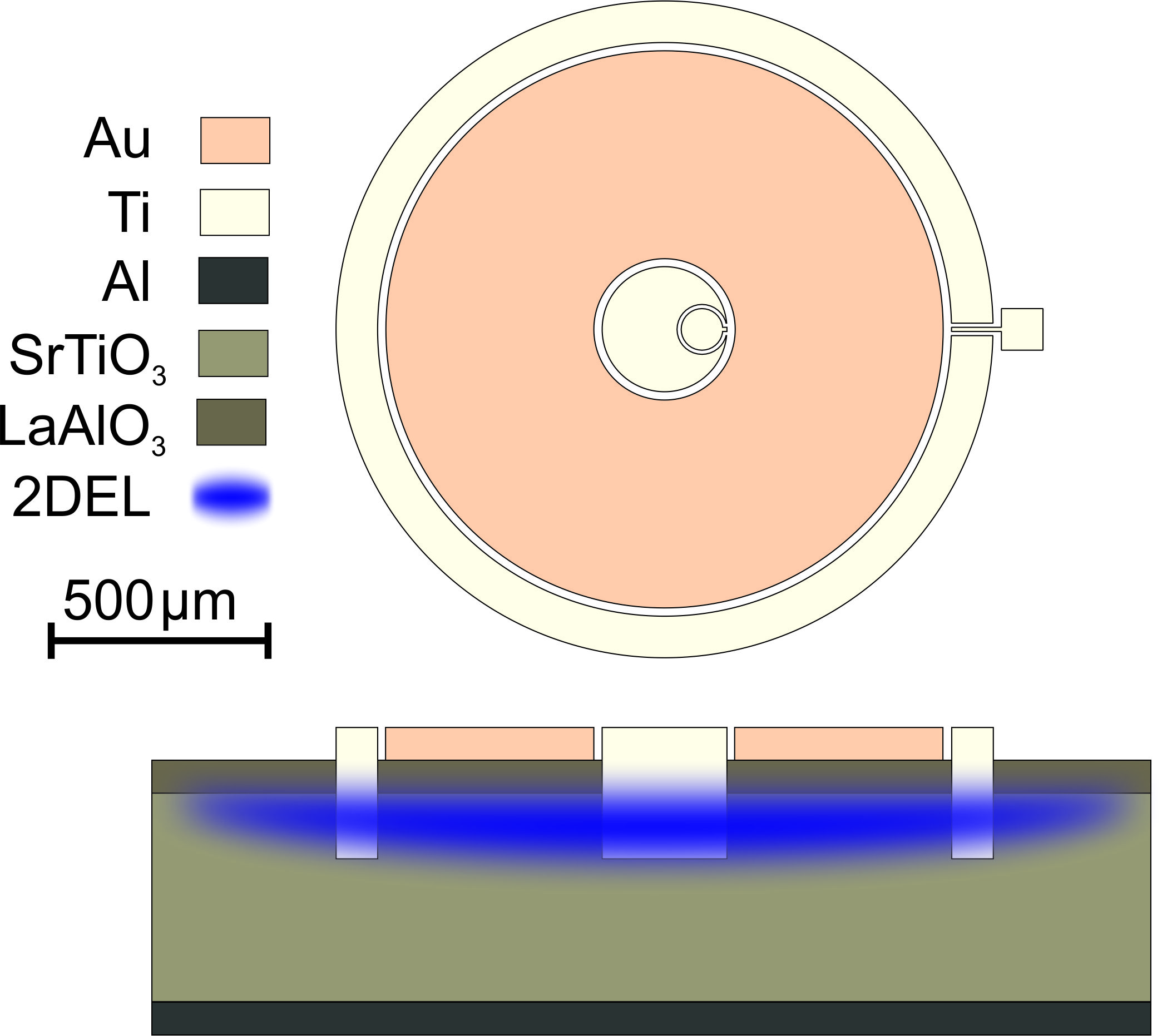}
}

\caption{Sample design of sample A. (a) Micrograph of a sample. (b) Schematic of a single device. Cross-sectional thicknesses are not to scale.}
\label{fig:device_layout}
\end{figure} 

Our sample design, shown in Fig. \ref{fig:device_layout}, comprises a gold top electrode for tunneling measurements and titanium contacts to the \LAOSTO interface 2DEL for in-plane measurements. The outer titanium electrode is grounded during measurement and acts as shielding. The surface area of the gold electrode equals $\approx$ \SI{1}{\square \milli \meter }. This large size was chosen to yield a sizable tunneling current with sufficient energy resolution to resolve the \si{\micro \volt}-gap. The \LAO layer of 4 unit cells serves a dual purpose. It acts both as the generator of the conducting interface and as the tunneling barrier. The samples were glued with silver paste to an aluminum bottom electrode prior to the measurements, which allows us to tune the carrier density of the sample using the high-$\kappa$ \STO substrate.  Descriptions of these  samples have already been published in \cite{Richter2013} and their suitability to measure both the superconducting gap and the phonon peaks in inelastic tunneling \cite{Boschker2015} has been demonstrated. 

\begin{figure}
\centering
\includegraphics[width=0.4\textwidth]{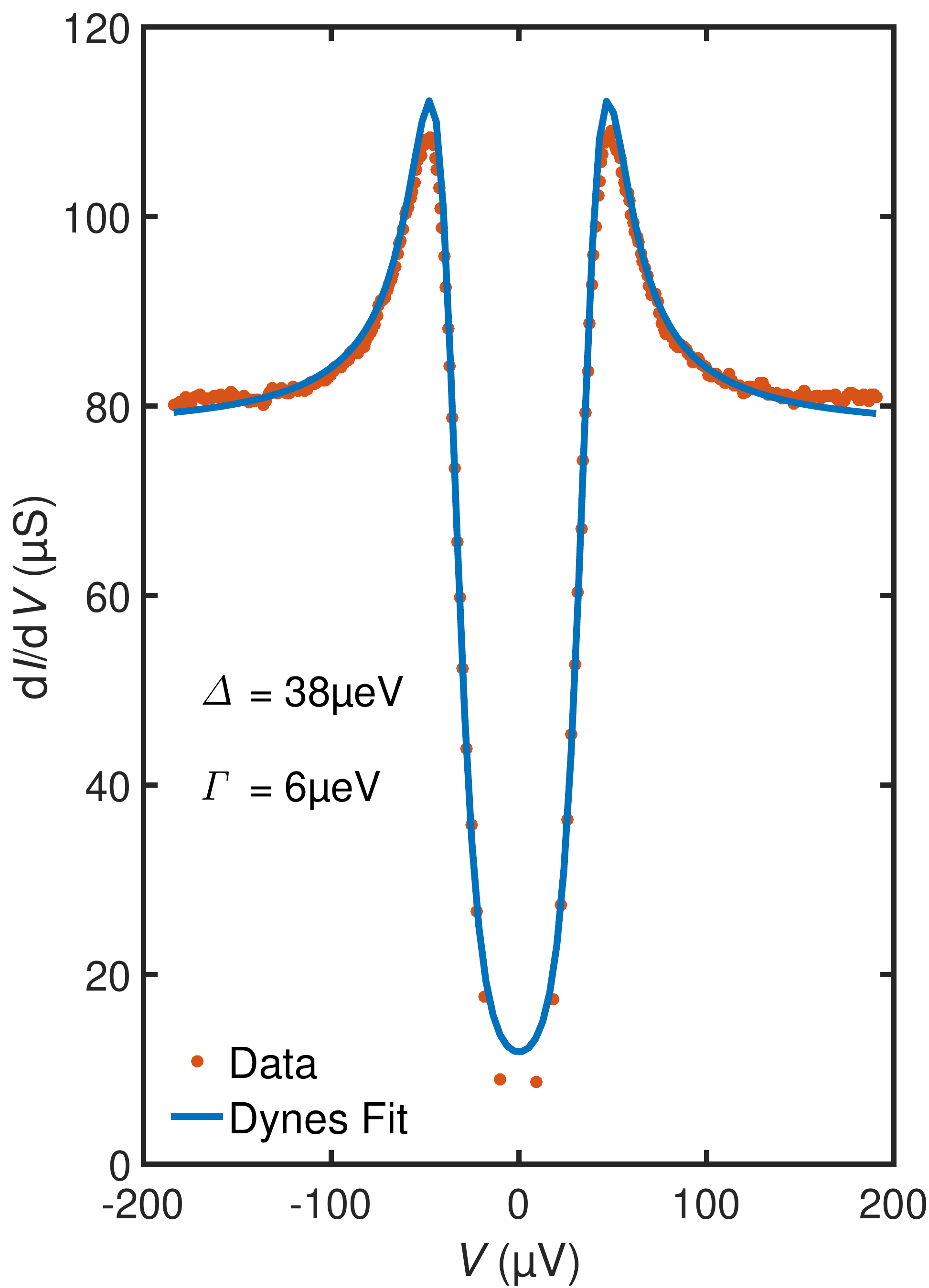}
\caption{A typical tunneling spectrum of sample A from the first measurement run without in-gap features measured at base temperature of the cryostat (\SI{50}{\milli \kelvin}). The Dynes fit\cite{Dynes1978} used to extract the gap width $\mathit{\Delta}$ and quasiparticle lifetime parameter $\mathit{\Gamma}$ is shown in addition to the data.}
\label{fig:no_features}
\end{figure} 

The \LAO layer was grown by pulsed laser deposition (PLD) onto $\mathrm{TiO_2}$-terminated \STO substrates with a fluence of \SI{0.8}{\joule \per \centi \meter \squared} at an oxygen pressure of \SI{8E-5}{\milli \bar}. Growth was monitored using reflectivity high-energy electron diffraction (RHEED).  Clear oscillations of the RHEED intensity indicated layer-by-layer deposition. Samples were subsequently annealed at \SI{400}{\milli \bar} oxygen pressure for one hour at \SI{600}{\celsius} and another hour at \SI{400}{\celsius} to ensure oxygen stoichiometry. A gold top layer was deposited \emph{in situ} to avoid adsorbates on the surface. Devices were structured on the sample surface by photolithography. Then, the gold tunneling electrode was defined by etching with KI+I solution and  electrodes to the 2DEL were created by  ion milling with subsequent electron beam evaporation of Ti.

Tunneling measurements were performed in 4-point configuration using a Keithley 6430 Femtoamp Sourcemeter as current source and  a Keithley 2001 or Keithley 2812 nanovolt meter. The polarity was such that a positive  bias corresponds to electrons tunneling from the 2DEL into the gold electrode. 

\captionsetup[subfloat]{width=6cm}

\begin{figure*}

\subfloat[]{
\includegraphics[width=0.45\textwidth]{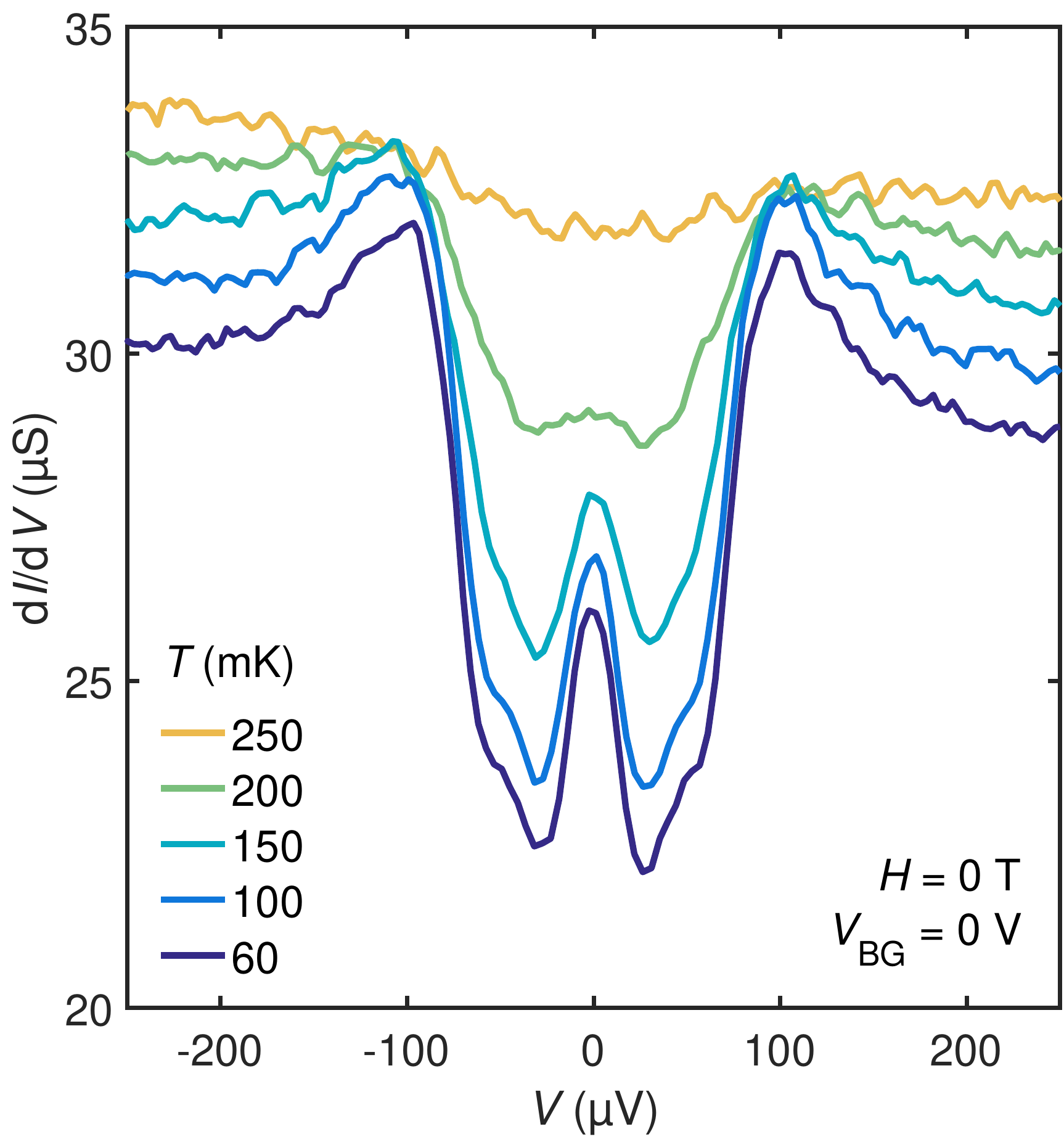}
}
\hfill
\subfloat[]{
\includegraphics[width=0.45\textwidth]{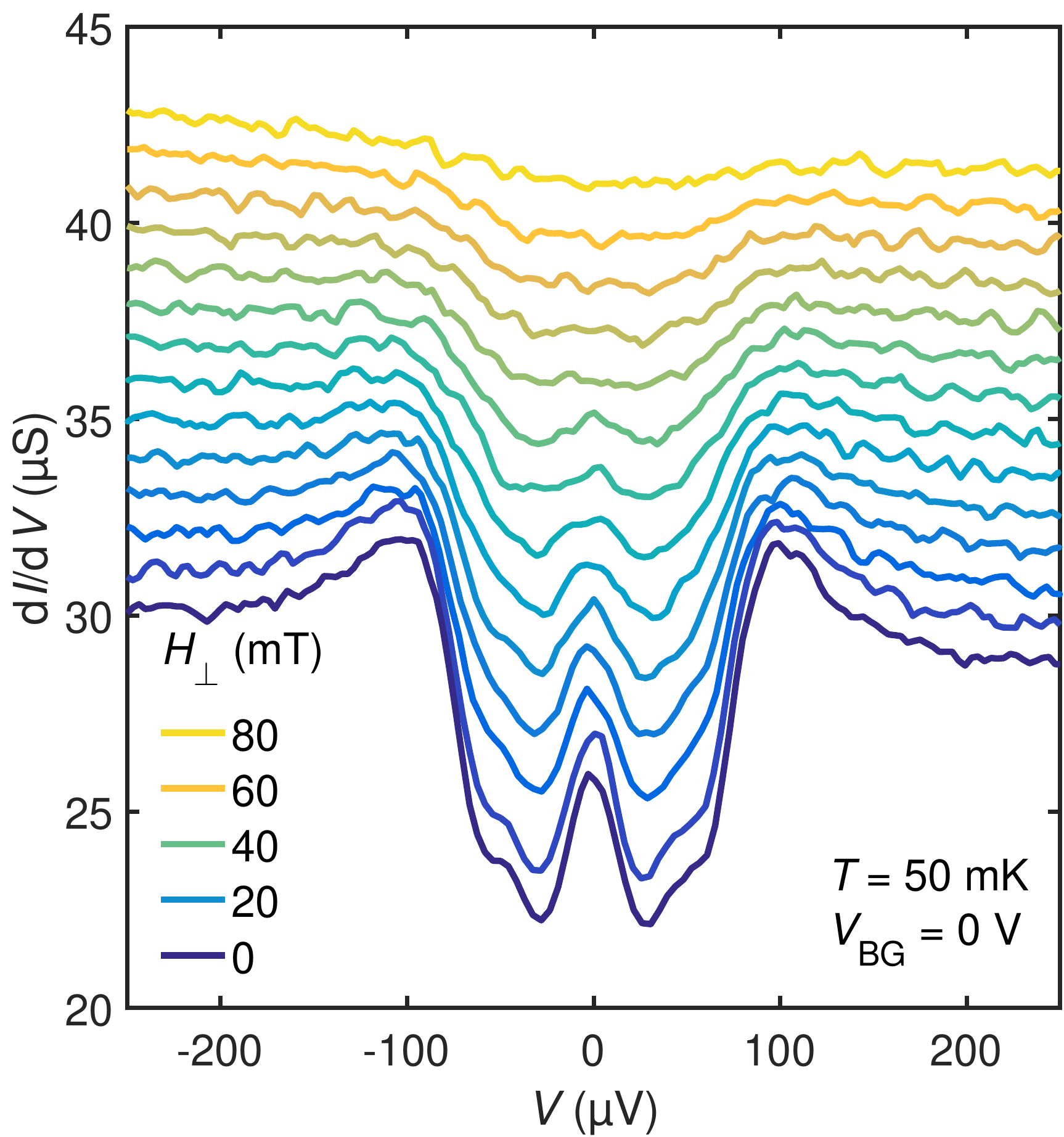}
}

\subfloat[]{
\includegraphics[width=0.45\textwidth]{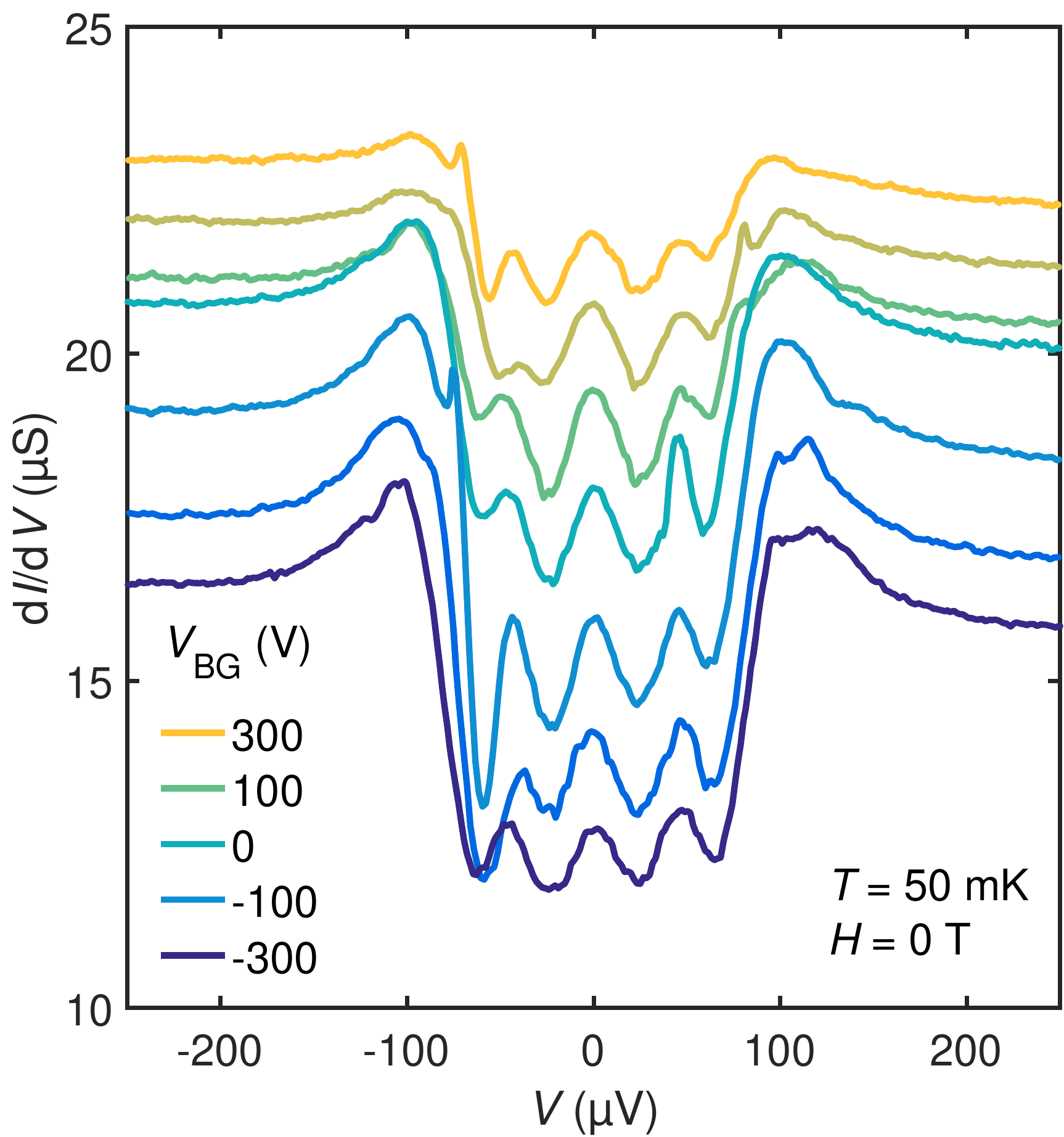}
}
\hfill \hfill \hfill \hfill
\subfloat[]{
\includegraphics[width=0.47\textwidth]{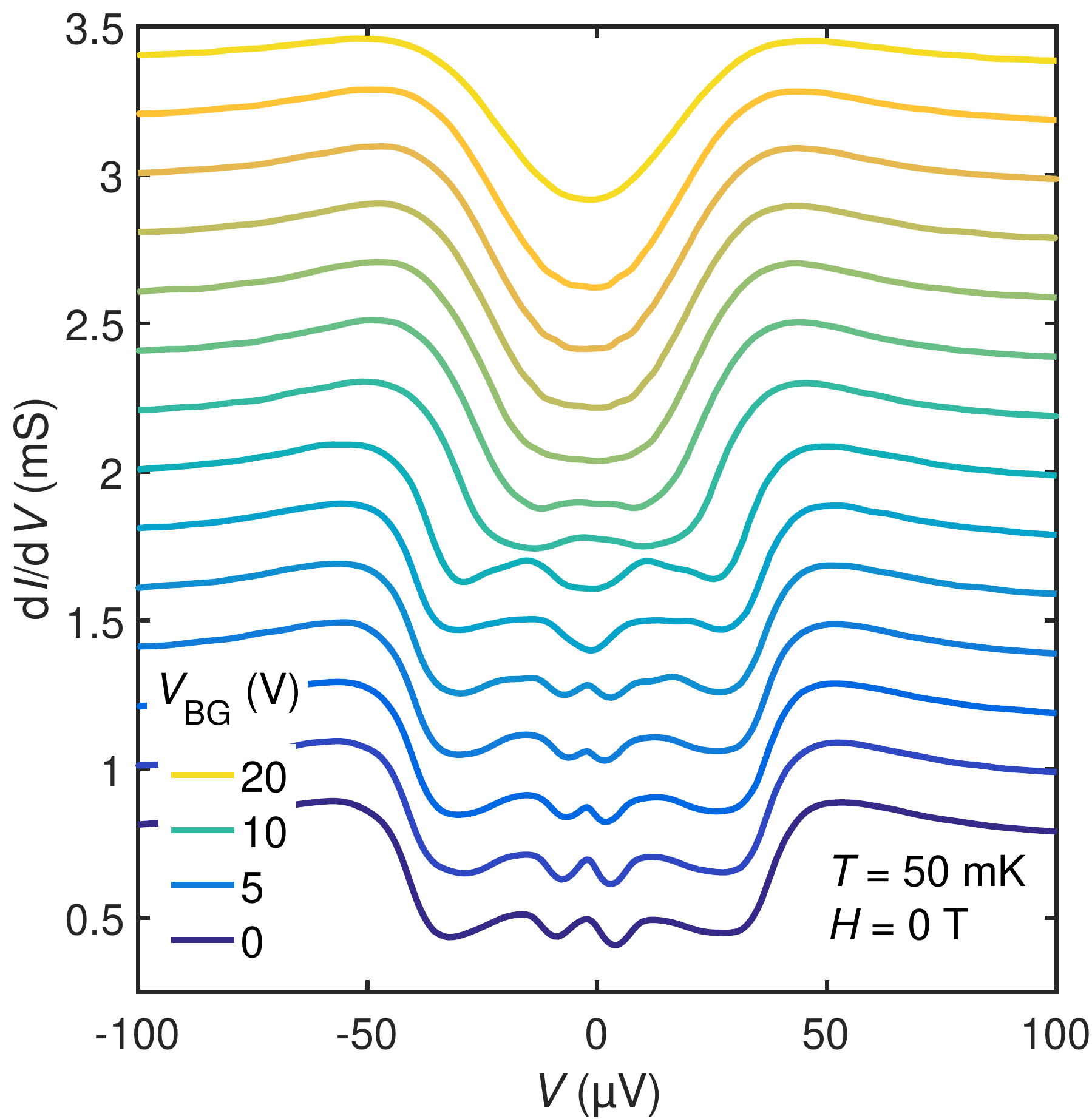}
}

\caption[in-gap Features]{Tunneling spectra of samples A and B illustrating the emergence of in-gap features: (a) Evolution of in-gap features in sample A with temperature (b) Evolution of in-gap features in sample A with magnetic field (c) Evolution of in-gap features in sample A with back-gate voltage, data measured in a different measurement run than (a) and (b). (d) Evolution of in-gap features in sample B with backgate voltage. In-gap features in (a) and (b) consist of a strong peak at zero bias and smaller peaks on either side inside the superconducting gap. In-gap features in (c) and (d) appear less systematic, but can be seen to be qualitatively similar to those of the other measurements when a different distribution of spectral weight between central and side peaks  is assumed. Curves are vertically shifted for visibility by \SI{1}{\micro \siemens} for (a)-(c) and by \SI{200}{\micro \siemens} for  (d), respectively. The conductivity is calculated by numerical differentiation with adaptive smoothing which does not change the in-gap feature characteristics. Negative backgate voltage corresponds to depletion of the interface, whereas positive backgate voltage corresponds to carrier accumulation.}
\label{fig:in-gap_states}
\end{figure*}

\captionsetup[subfloat]{width=0cm}

The shape of the superconducting gap observed in the tunneling spectra of \LAOSTO -interfaces follows the prediction for a standard  s-wave BCS superconductor taking into account finite quasiparticle lifetime \cite{Dynes1978}. However, in some samples and in some measurement runs, we observed distinct in-gap features inside the superconducting gap. These features are observable regardless of the sweep direction or sweep rate of the measurement. The in-gap features appear or disappear  between different measurement runs, i.e. after a thermal cycle to room temperature: On sample A, in the first measurement run, standard tunneling spectra were observed on both of the devices  which had been bonded (Fig. \ref{fig:no_features}). After a thermal cycle to room temperature, both devices   showed in-gap features such as shown in Fig. \ref{fig:in-gap_states} (a) and (b). In a number of subsequent warming and cooling cycles, in-gap features of varying strength were observed in this sample (Fig. \ref{fig:in-gap_states} (c)). On sample B, in the first measurement run the in-gap features shown in Fig. \ref{fig:in-gap_states} (d) were observed after saturating the sample with charge carriers at high positive backgate voltage and then returning to zero backgate. The in-gap features were not observed in subsequent measurement runs on sample B. It is not clear  why only specific samples show these anomalies and others do not.

\section{ Results}

The in-gap structures are shown in Fig. \ref{fig:in-gap_states} for two different samples grown in two different PLD systems. Fig. \ref{fig:in-gap_states} (a) shows tunnel spectra of sample A as a function of temperature, Fig. \ref{fig:in-gap_states} (b) shows spectra of sample A as a function of magnetic field, Fig. \ref{fig:in-gap_states} (c) shows spectra of sample A from a different measurement run as a function of back-gate voltage and Fig. \ref{fig:in-gap_states} (d) shows spectra from sample B as a function of back-gate voltage. The in-gap-features observed in   Fig. \ref{fig:in-gap_states} (a) and (b) consist of a strong peak at zero bias and two smaller peaks, one at either side of the gap. Additionally, the width of the gap is increased compared to the standard spectra with the smaller peaks appearing at voltage values comparable to those of the coherence peaks in the standard spectra. The in-gap features disappear at the same temperature and field scales as the superconducting gap itself, i.e. it is neither possible to observe the features without the gap, nor the gap without  the features. Both the application of field and temperature suppress the gap and the features, but do not destroy them. Temperature was increased to \SI{1}{\kelvin} and field to \SI{5}{\tesla}, after which the features reappeared when returning to base temperature and zero field. The intrinsic charge carrier density of sample A was so high that the 2DEL could not be depleted completely with the gate voltages accessible in our experiment. Only minor changes of the in-gap states were observed for -300 $<$ $V_\textrm{G}$ $<$ 300 V (Fig. \ref{fig:in-gap_states}(c)).

In contrast to sample A, the spectra of sample B depend strongly on the applied gate voltage. For $V_\textrm{G}$ $>$ 10 V, no in-gap states were observed. For $V_\textrm{G}$ $<$ 10 V, first a single zero bias peak is present, then for decreasing $V_\textrm{G}$ gradually more peaks appear. At first sight, the tunneling spectra observed in Fig. \ref{fig:in-gap_states} (c) and (d) appear to be quantitatively different from those observed in Fig. \ref{fig:in-gap_states} (a) and (b). However, their similarity becomes obvious when a different distribution of spectral weight between the central and the side peaks is taken into account. 

\begin{figure}

\includegraphics[width=0.4\textwidth]{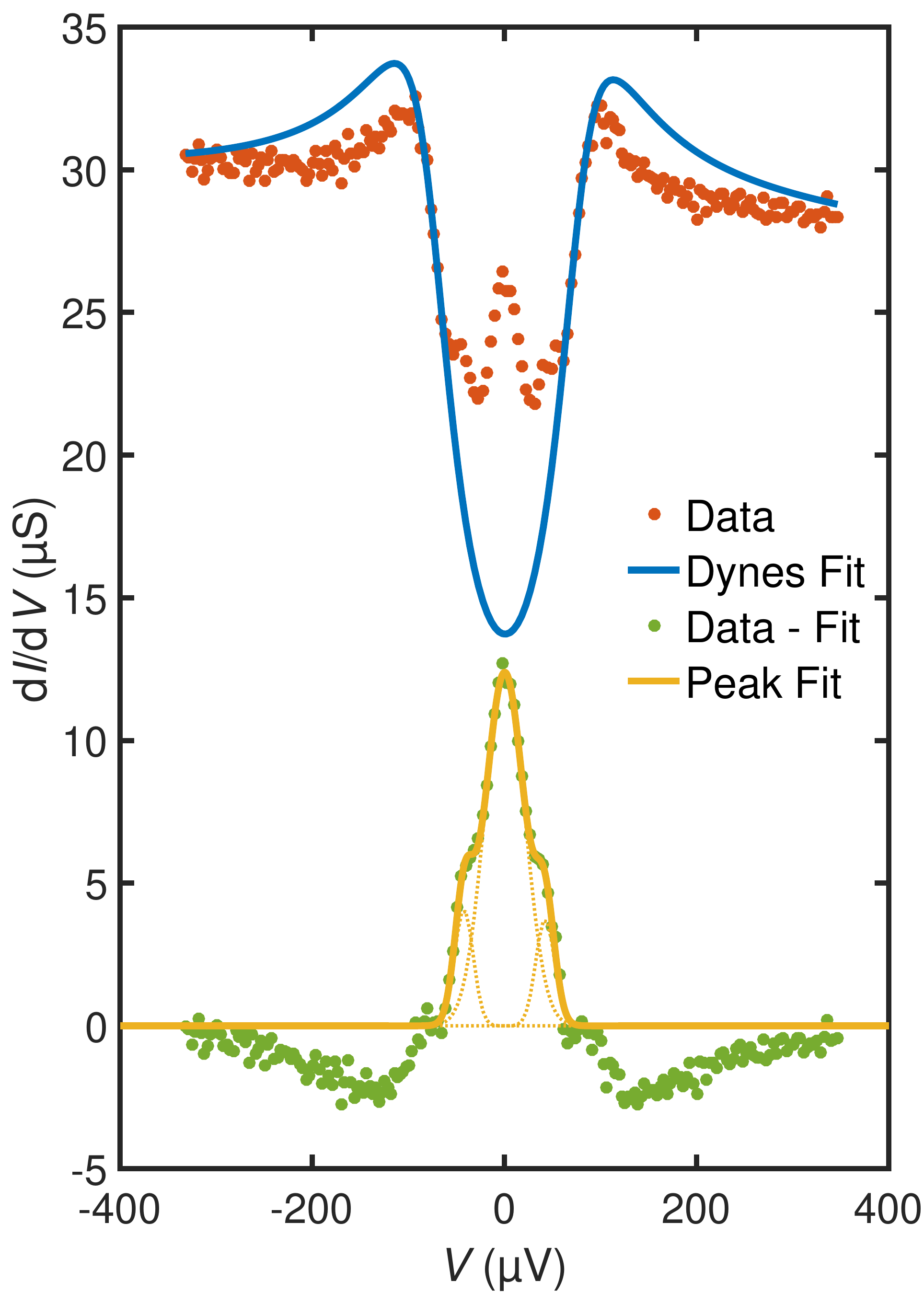}

\caption[]{Illustration of the quantitative analysis used to characterize the in-gap peaks: A Dynes fit to the gap part without peaks is used to calculate the difference between expected and measured conductivity values.   Three Gaussians are fitted to the subtracted peak to determine  position, height and FWHM of the central peak and the separation of the side peaks. The dotted yellow lines indicate the three single Gaussians and the continuous yellow line indicates the sum of all three Gaussians.  Note that because spectral weight is shifted from the coherence peaks into the in-gap peaks, the coherence peaks are not described well by the lifetime-broadened BCS fit.}
\label{fig:peak_analysis}
\end{figure}

To gain quantitative information on the properties of the observed in-gap-peaks, we performed the fitting routine illustrated in Fig. \ref{fig:peak_analysis}: For each curve, a lifetime-broadened BCS fit (Dynes fit) was adjusted to the part of the superconducting gap without peaks to create a reference curve corresponding to a standard \LAOSTO gap. Since spectral weight is shifted from the coherence peaks into the in-gap peaks, the coherence peaks are not described well by this fit. The Dynes curve was subtracted from the data points to obtain the deviation from standard superconducting behavior. Three Gaussians were fitted to this subtracted peak to obtain the position, size and full width at half maximum (FWHM) of the central zero-bias peak and the separation of the side peaks.

\begin{figure*}

\subfloat[]{
\includegraphics[width=0.45\textwidth]{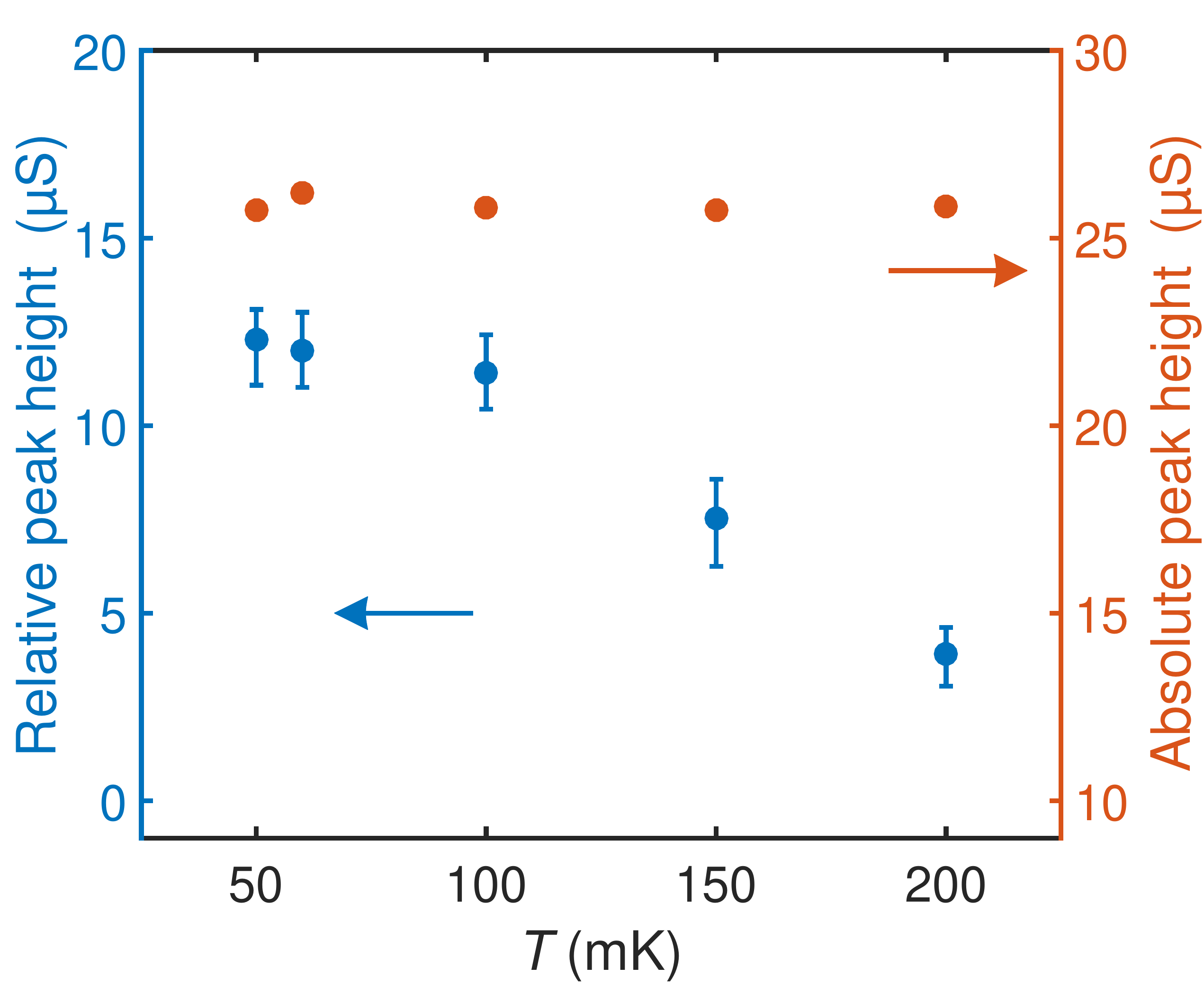}
}
\hfill
\subfloat[]{
\includegraphics[width=0.45\textwidth]{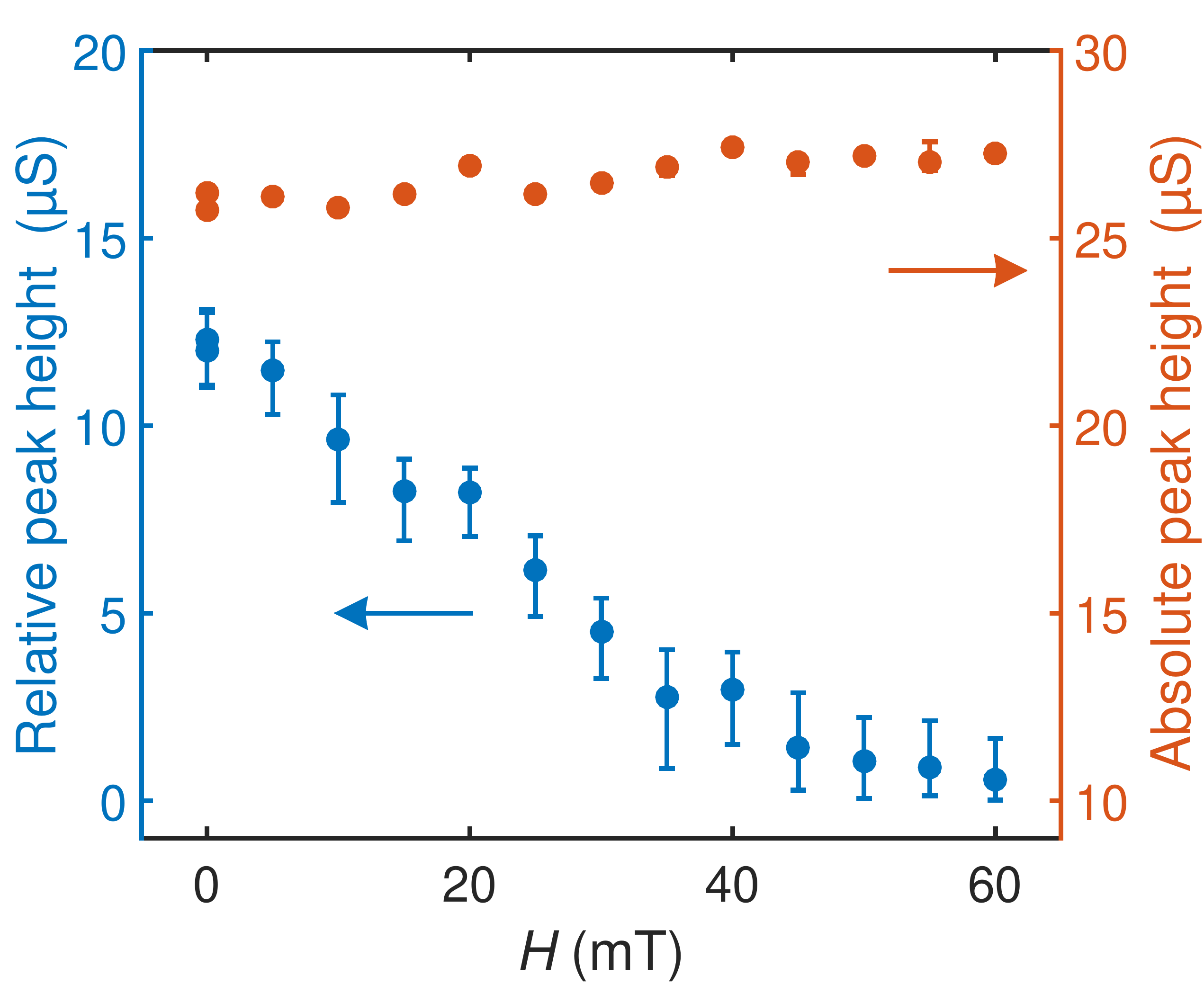}
}

\subfloat[]{
\includegraphics[width=0.45\textwidth]{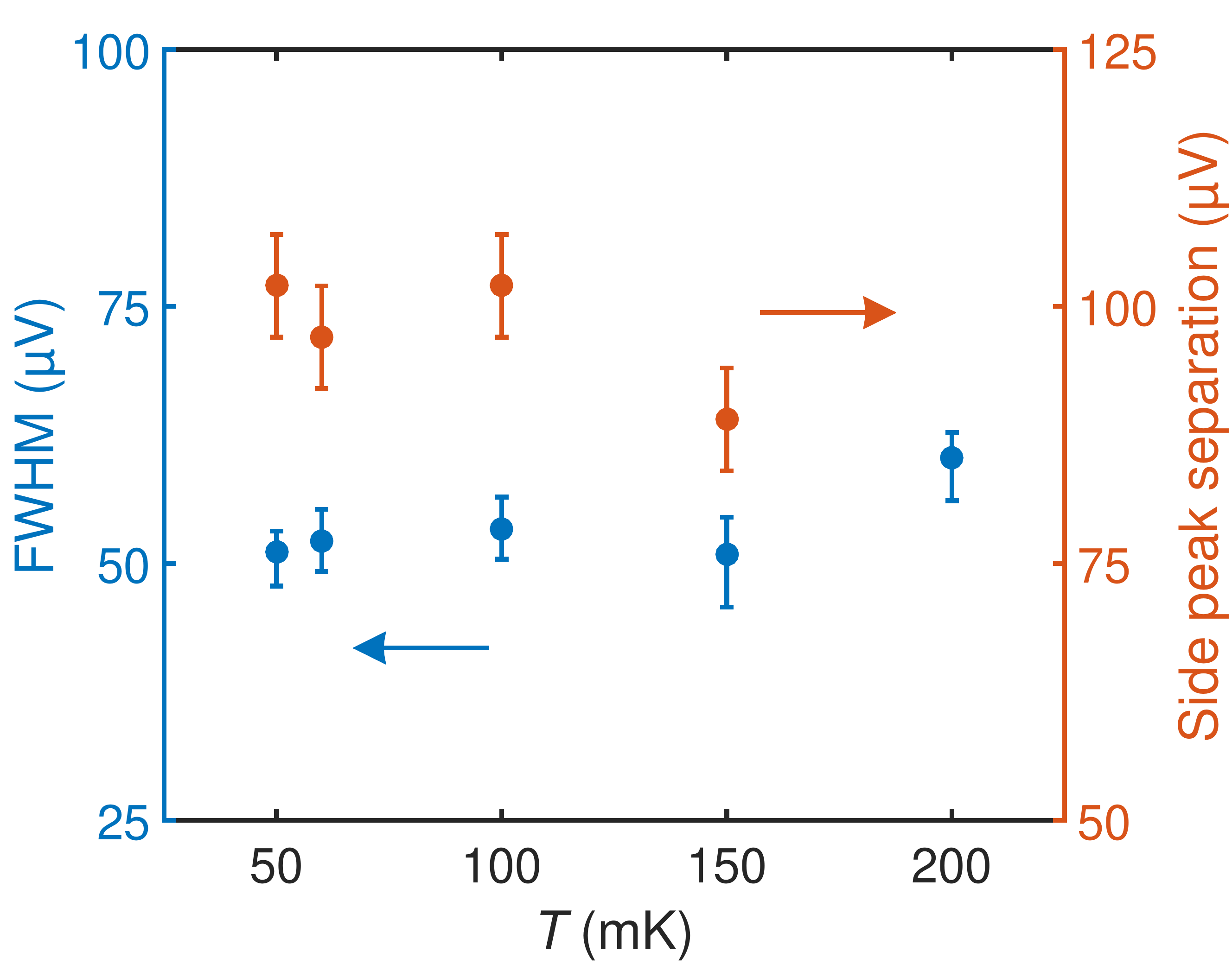}
}
\hfill
\subfloat[]{
\includegraphics[width=0.45\textwidth]{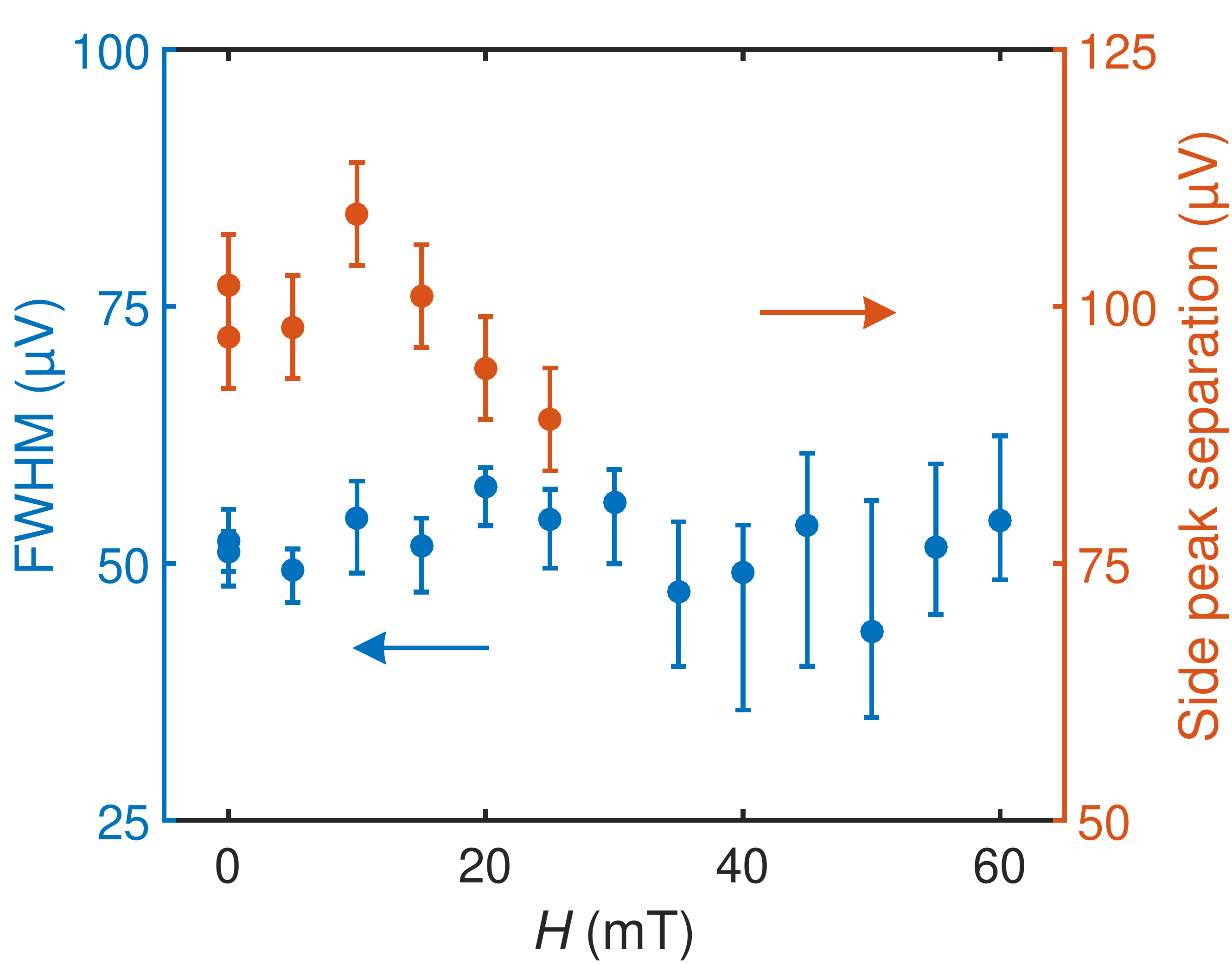}
}

\caption[]{Evolution of height and FWHM of the zero bias peak and of the separation of the side peaks observed in sample A. In (a) and (b)  we plot both the relative height of the peak, i.e. the difference between data and BCS fit (blue), and the absolute size of the peak in the raw data (red). Error bars indicate values obtained for BCS fits where the sum of weighted residuals is double that of the optimal fit. Since the absolute height of the peak depends only marginally on the fit, error bars for the absolute peak are smaller than the data points (a) Evolution  with temperature. (b) Evolution  with magnetic field. Whereas the relative peak height decreases with increasing magnetic field or temperature, the absolute peak height remains almost constant.
In (c) and (d) we plot the evolution of the FHWM of the central peak (blue) and the separation of the side peaks (red). Again, error bars are obtained from data calculated for BCS fits with twice the minimum residual. (c) Evolution  with temperature. (d) Evolution  with magnetic field. Both the FWHM of the central peak and the spacing of the side peaks remain approximately constant over the  field and temperature range in which the superconducting gap can be observed. Values of the side peak separation are only shown for those curves in which side peaks can be clearly discerned.}
\label{fig:peak_evolution}
\end{figure*}

Results of this analysis are shown in Fig. \ref{fig:peak_evolution}: The size of the subtracted peak decreases monotonically with increasing temperature or magnetic field and disappears at the same values of temperature and field at which the gap closes. On the other hand, the absolute height of the peak remains almost constant over the entire measurement range close to the value corresponding to  $\mathrm{d}I / \mathrm{d}V$ outside the gap. Also both the FWHM of the central peak and the separation of the outer peaks is independent of temperature and magnetic field. Since only few data points are available for the Dynes fits, the error bars given in Fig.  \ref{fig:peak_evolution} are based on the fit uncertainty: For the optimal fit, the sum of residuals, i.e. of the weighted difference squares between data and fit curve, is minimal. Error bars denote results from fits at the border of the confidence interval, for which the sum of residuals is twice as large as for the optimal fit.

\section{Discussion}

\subsection*{Origin of the in-gap states}

We now discuss the applicability of the phenomena described in Section II to our experimental data. Since we always observe either three peaks or no peaks at all, we conjecture that all peaks are caused by the same mechanism. Therefore, we concentrate our discussion on those mechanisms which  can account for all observed peaks.

\subsubsection*{Kondo resonance}
A strong peak at zero bias can be caused by a Kondo resonance. However, the observed side peaks cannot be easily explained in a Kondo framework. Also, a zero-bias-anomaly caused by Kondo scattering is expected to split with increasing magnetic field, which we do not observe. Finally,  the zero-bias-peak in the measurements disappears at the same temperature and field as the superconducting gap, hinting at an intimate connection between the peak and superconductivity. We therefore conclude that Kondo scattering is unlikely to be the origin of the observed in-gap states.

\subsubsection*{Anderson-Appelbaum states}
Similarly to Kondo resonances, Anderson-Appelbaum states are not connected to superconductivity. Also, the side-peak separation does not vary linearly with external magnetic field, as expected for these states. We therefore do not consider them as likely candidates for the origin of the observed in-gap states.

\subsubsection*{Barrier and surface impurities}
Both the barrier and surface impurity models account only for features at zero bias but cannot explain the side peaks at finite voltage. We therefore conclude that they are unlikely to be the origin of the observed in-gap states.

\subsubsection*{Superconductor-insulator-superconductor (SIS) tunneling}

Both the zero bias current and the broadening of the gap to approximately twice the value of standard \LAOSTO (cf. Fig. \ref{fig:no_features}) can clearly be seen in the data from sample A. Thus the tunnel characteristics observed here strongly resemble that of a hypothetical SIS junction, with the side peaks representing the coherence peaks of the inner gap. However, there is only one superconducting electrode in the tunnel junction. It is implausible that the superconducting \LAOSTO interface  induces superconductivity in the gold top electrode through the insulating \LAO tunneling barrier. We therefore conclude that SIS tunneling is an unlikely explanation of the observed in-gap states.

\subsubsection*{Multiband superconductivity}
The side peaks observed in our tunneling spectra are explicable as the coherence peaks of a second superconducting band. However, the strong peak at zero bias cannot be explained in this framework. Also, the coherence peaks of a second gap should move closer together as the gap closes with increasing magnetic field or temperature, which is not observed. Therefore, we conclude that multiband superconductivity is not a likely explanation of our observations.

\subsubsection*{Caroli-de Gennes-Matricon states}

In our experiment, the peak at zero bias becomes weaker with increasing magnetic field, i.e., with an increasing number of vortices. Thus it shows the opposite behavior  of that expected for CdGM-states. We therefore believe that the CdGM-mechanism is not the origin of the observed in-gap states.

\subsubsection*{Yu-Shiba-Rusinov states}

YSR states can cause multiple peaks inside the gap of superconductors with magnetic impurities. However, YSR states always appear in pairs around zero, therefore the single peak at zero bias  shown in Fig. \ref{fig:in-gap_states} would be explicable only as a smeared pair of two YSR states. Such a smeared peak should broaden as the two constituent peaks move apart with increasing magnetic field, which is not observed. Also the two peaks on either side remain at the same position independent of magnetic field, in contrast to the outward movement expected for YSR states.  The range of magnetic fields accessible in our experiment is limited because the critical magnetic field of our \LAOSTO samples is small. We therefore have performed simulations to assure that the movement of the peak positions should indeed be observable on the magnetic field scales investigated, if the peaks were due to YSR states. Since we observe neither a broadening of the central peak nor a movement of the side peaks, we conclude that YSR states are most likely not the origin of the observed in-gap states.

\subsubsection*{Majorana bound states}

The zero bias peak observed on sample A is of almost constant height as expected for a MBS. The fixed position of the peak at zero bias is consistent with the hypothesis of MBS, with the peaks at finite voltage bias possibly indicating Majoranas with overlapping wavefunctions \cite{Nilsson2008, Beenakker2013, Flensberg2010}. The \LAOSTO -interface has been suggested as a candidate host for MBS~\cite{Mohanta2014}, since it comprises the basic ingredients required for topological superconductivity \textit{viz.} $s$-wave superconductivity, Rashba spin-orbit coupling and magnetism. The appearance of Majorana modes usually requires specific configurations of the 2DES, e.g., magnetic oxygen vacancies arranged in a linear chain which is quite unlikely to be readily available at the inhomogenous interface.  However, scanning superconducting quantum interference device (SQUID) measurements provide evidence for the existence of one-dimensional conducting channels at the domain boundaries of substrate SrTiO$_3$~\cite{Kalisky2013}.  The possibility of MBS as the origin of the in-gap states cannot be ruled out.

\subsubsection*{Andreev bound states}

ABS can generate conductance peaks  at zero bias as well as  at finite energy, consistent with our observations. However, a zero-bias conductance peak caused by an ABS is expected to split upon increasing the magnetic field, which we do not observe. Andreev bound states usually occur at junctions of superconductors with metals which have sufficiently high conductivity, whereas in our samples, gold electrode and superconductor are separated by an insulating layer. 
However, the NS contact can be situated within the 2DES, either as an in-plane combination of normal and superconducting islands or as separate normal and superconducting layers. The application of gate voltage changes the carrier density and hence the superconducting volume fraction, changing the number of the ABS.

\subsubsection*{Odd-frequency spin-triplet pairing}
Odd-frequency spin-triplet pairing can account for both  the zero bias peak and the side peaks. It requires an inhomogeneous magnetization at the interface, which in some regions of the sample generates zero-bias peaks and peaks at finite bias in other regions. Since the area of our tunnel junctions is larger than the domain size, the peaks from different regions are observed together in our spectra. However, the gap size for singlet pairing and triplet pairing is likely to differ and therefore an averaging over triplet and singlet regions should show a double gap with two pairs of coherence peaks, unless the condensate is always either completely singlet or completely triplet.   Magnetism has been observed to coexist with superconductivity in \LAOSTO \cite{Li2011, Bert2011} and it has been shown to be rather superparamagnetic than truly ferromagnetic in nature  \cite{Li2011}. Therefore, the difference between the superparamagnetic domains could generate an inhomogeneous magnetization if the variation of magnetization between domains is strong enough.

\subsection*{Origin of the dependence on thermal cycling}

Finally, we speculate on some scenarios that could explain the fact that the in-gap states are only observed in a small fraction of samples and that they depend on thermal cycling. A tentative explanation for the appearance and disappearance of sub-gap states may lie in the domain structure of the \STO substrate: as the crystal structure changes from cubic to tetragonal when the crystal is cooled below 105 K, a domain structure forms, which is randomly different in each measurement run. If we assume the in-gap states to depend on a specific domain configuration, then this scenario can also explain why the in-gap features sometimes disappear irreversibly when sweeping the gate voltage, because the gate voltage influences the \STO domain structure \cite{Honig2013}. Another possible explanation is that the in-gap states depend on a specific configuration of defects in the samples, for example oxygen vacancies. Oxygen vacancies form magnetic centers\cite{Pavlenko2012} and could thereby influence the superconductivity. Finally, we mention that the back-gate voltage also affects the thickness $d$ of the superconducting sheet, with $d$ increased up to a factor of three in the overdoped region \cite{Gariglio2016}.

\section{Conclusions}
Using tunneling spectroscopy on the two-dimensional superconductor, we have observed in-gap features in spectra of several superconducting \LAOSTO tunnel devices. The features appear and disappear non-deterministically between different warming and cooling cycles. The in-gap states were found not to move with either temperature or magnetic field, yet to change under the application of back-gate voltage. The real challenge is to disentangle the true origin of these in-gap states with the limited information available at the buried interfaces. None of the known mechanisms that can cause in-gap states easily explains the results. Conceivable scenarios involve Majorana bound states, Andreev bound states, or the presence of an odd-frequency spin-triplet component in the superconducting order parameter caused by an inhomogeneous ferromagnetic state in the electron system. 

\begin{acknowledgements}
We thank Y. Fominov, C. Renner, and A. Santander-Syro for enlightening discussions.
\end{acknowledgements}

\bibliographystyle{apsrev4-1}
\bibliography{bibliography}

\end{document}